\documentclass[sigconf]{acmart}

\usepackage{microtype}
\usepackage{booktabs}
\usepackage{amsmath}
\usepackage{amsthm}
\usepackage{graphicx}
\usepackage{wrapfig}
\usepackage{dblfloatfix}
\usepackage{algorithm}
\usepackage{algpseudocode}

\setcopyright{none}
\acmConference[OARS '26]{Workshop on Online and Adaptive Recommender Systems}{September 28, 2026}{Minneapolis, MN, USA}
\acmDOI{}
\acmISBN{}

\title{MISO: Model-Internal-State-Guided Optimization for Ranking Models}

  \author{Yongzhe Zhang\texorpdfstring{\textsuperscript{*}}{*}} \email{yongzhe@meta.com} \affiliation{\institution{Meta Inc}\country{United States}}
\author{Xiaoyu Deng\texorpdfstring{\textsuperscript{*}}{*}} \email{xiaoyud@meta.com} \affiliation{\institution{Meta Inc}\country{United States}}
\author{Yifan He\texorpdfstring{\textsuperscript{*}}{*}} \email{yifanhe1999@meta.com} \affiliation{\institution{Meta Inc}\country{United States}}
\author{Mengying Sun\texorpdfstring{\textsuperscript{*}}{*}} \email{mengyingsun@meta.com} \affiliation{\institution{Meta Inc}\country{United States}}
\author{Yijia Liu} \email{yijialiu@meta.com} \affiliation{\institution{Meta Inc}\country{United States}} 
\author{Sheng Luo} \email{shengluo@meta.com} \affiliation{\institution{Meta Inc}\country{United States}} 
\author{Marcio Porto} \email{mflporto@meta.com} \affiliation{\institution{Meta Inc}\country{United States}} 
\author{Anish Khazane} \email{akhazane@meta.com} \affiliation{\institution{Meta Inc}\country{United States}} 
\author{Zhuo Li} \email{zlli@meta.com} \affiliation{\institution{Meta Inc}\country{United States}} 
\author{Huiping Yao} \email{huiping@meta.com} \affiliation{\institution{Meta Inc}\country{United States}} 
\author{Swathi Hrishikesh} \email{swathih@meta.com} \affiliation{\institution{Meta Inc}\country{United States}} 
\author{Jing Chen} \email{jingchenfb@meta.com} \affiliation{\institution{Meta Inc}\country{United States}} 
\author{Dennis Choi} \email{dennchoi@meta.com} \affiliation{\institution{Meta Inc}\country{United States}} 
\author{Steven Liu} \email{stevenliu0305@meta.com} \affiliation{\institution{Meta Inc}\country{United States}} 
\author{Xiaoya Wang} \email{xiaoya213@meta.com} \affiliation{\institution{Meta Inc}\country{United States}} 
\author{Emmy Wang} \email{emmyw@meta.com} \affiliation{\institution{Meta Inc}\country{United States}} 
\author{Kangfu Zheng} \email{zhengkf02@meta.com} \affiliation{\institution{Meta Inc}\country{United States}} 
\author{Xingyuan Wang} \email{sylviawang@meta.com} \affiliation{\institution{Meta Inc}\country{United States}} 
\author{Zhiwen Chen}
\email{kylechen@meta.com}
\affiliation{%
  \institution{Meta Inc}
  \country{United States}
}

\author{Yang Jin}
\email{yangjin@meta.com}
\affiliation{%
  \institution{Meta Inc}
  \country{United States}
}

\author{Jiang Liu}
\email{jiangliu@meta.com}
\affiliation{%
  \institution{Meta Inc}
  \country{United States}
}

\author{Haoyu Zhou}
\email{haoyuzhou@meta.com}
\affiliation{%
  \institution{Meta Inc}
  \country{United States}
}
\author{Lexi Luo} \email{lexiluo@meta.com} \affiliation{\institution{Meta Inc}\country{United States}} 
\author{Keyi Chen} \email{keyic@meta.com} \affiliation{\institution{Meta Inc}\country{United States}} 
\author{Hao Yan} \email{yanh@meta.com} \affiliation{\institution{Meta Inc}\country{United States}} 
\author{Peggy Yao} \email{peggy@meta.com} \affiliation{\institution{Meta Inc}\country{United States}} 
\author{Alireza Vahdatpour} \email{alirezav@meta.com} \affiliation{\institution{Meta Inc}\country{United States}}
\author{Santanu Kolay} \email{skolay@meta.com} \affiliation{\institution{Meta Inc}\country{United States}} 
\author{Yi Meng} \email{ymeng@meta.com} \affiliation{\institution{Meta Inc}\country{United States}} 
\author{Bilal Fadlallah} \email{bhf@meta.com} \affiliation{\institution{Meta Inc}\country{United States}} 
\author{Gursharan Singh} \email{gurshi@meta.com} \affiliation{\institution{Meta Inc}\country{United States}} 
\author{Prabhakar Goyal} \email{prgoyal@meta.com} \affiliation{\institution{Meta Inc}\country{United States}} 

\hypersetup{
  pdftitle={MISO: Internal-State-Guided Optimization for Ranking Models},
  pdfsubject={RecSys workshop submission},
  pdfkeywords={recommender systems, AutoML, model optimization, internal states, ranking models}
}

\begin{document}

\begin{abstract}
Ranking models are repeatedly refined within established model families, yet the choice of which component to scale, replace, or retire is often guided by expensive trial-and-error. We present \textbf{Model Internal State Optimization (MISO)}, a systems workflow that uses model internal states (MIS)-including parameters, activations, gradients, and normalization statistics—to prioritize such local optimization decisions. MISO extracts MIS from a trained ranking model, aggregates them into ranking, alignment, and comparison signals, and converts those signals into a small set of interpretable candidate edits. Because MIS are re-extracted after each retraining cycle, MISO naturally supports an adaptive optimization workflow that tracks evolving model behavior as data distributions and system requirements shift over time. In an ads ranking case study, MISO improves normalized entropy while requiring substantially fewer validation runs than expert-driven and black-box scaling workflows, offering a practical middle ground between manual tuning and opaque automated search.
\end{abstract}


\maketitle \renewcommand{\thefootnote}{\fnsymbol{footnote}} \footnotetext[1]{Equal Contribution.} \renewcommand{\thefootnote}{\arabic{footnote}}

\section{Introduction}

Large-scale ranking and recommendation systems are rarely redesigned from scratch. More often, engineers repeatedly make local decisions within an established model family: where to add capacity, which module to replace, which feature pathway to retire, or which nearby variant merits another expensive training run. These decisions must balance ranking quality, latency, and operational cost, but are still commonly driven by manual inspection and trial-and-error.

Although trained models already expose a rich collection of internal signals—parameters, activations, gradients, normalization statistics, and intermediate representations—these signals are seldom used as a systematic optimization interface. End-to-end metrics reveal whether a variant improved, but do not directly indicate where a model is under-utilized, which representations are unstable, or why two nearby variants diverge.

Turning internal signals into reliable engineering actions is difficult. MIS are high-dimensional and heterogeneous, while the available actions are coarse-grained and structural. A useful system must therefore summarize low-level evidence into a small number of proposals that engineers can inspect, validate, and reject under a limited experimentation budget.

Existing workflows leave a gap between expert-driven tuning and black-box AutoML. The former can exploit domain knowledge but is difficult to reproduce; the latter can automate exploration but typically treats a model as an opaque object and may require many costly trials. Neither approach makes the model's own internal evidence a first-class input to the optimization loop.

We present \textbf{MISO}, a machine learning systems framework for internal-state-guided optimization of ads ranking models. MISO closes the loop between internal behavior and architectural actions: it extracts MIS, aggregates them into actionable summaries, and uses those summaries to prioritize a small set of candidate modifications for validation.

MISO provides (1) a unified abstraction and extraction interface for MIS, (2) ranking, alignment, and comparison primitives that turn fine-grained signals into decision-oriented summaries, and (3) a budgeted decision loop that validates only the most promising edits. It shares AutoML's aim of reducing manual effort, but targets repeated local refinement rather than broad architecture discovery.

We evaluate MISO in an ads ranking case study spanning several model scales and three recurring optimization tasks. The results show that MIS-guided proposals can improve ranking quality while reducing the number of training runs needed to identify a useful configuration.

\textbf{Contributions.} Our main contributions are:
\begin{itemize}
    \item We introduce \textbf{MISO}, a systems framework that treats model internal states as first-class optimization signals for local ranking-model refinement.
    \item We design three classes of MIS aggregation primitives (ranking, alignment, comparison) that bridge fine-grained signals to actionable architectural decisions.
    \item We demonstrate on ranking models that MISO achieves up to $2.5\times$ the relative NE improvement of expert-driven tuning while reducing exploration cost by 84--94\%.
\end{itemize}

\section{Problem Statement and Design Goals}

\subsection{Problem Statement}

We consider \emph{local model optimization guided by model internal states} (MIS). Given a trained ranking model $M$ with parameters $\theta$, a training or evaluation snapshot induces internal signals such as parameter tensors, activations, gradients, normalization statistics, and neuron-level importance measures. We denote this collection by $\mathcal{S}(M)$.

The objective is to use $\mathcal{S}(M)$ to prioritize architectural and capacity-related edits—such as selective scaling, pruning, module replacement, and feature reconfiguration—under a limited experimentation budget. Rather than searching a broad architecture space, MISO seeks a short sequence of validated modifications $\{\Delta M\}$ that improves the target ranking metric while respecting operational constraints such as latency and training cost.

This problem is difficult because MIS are high-dimensional, heterogeneous, and highly localized, whereas optimization actions are coarse-grained and structural. To bridge this gap, MISO is designed around the following goals: 

\subsection{Design Goals}

\textbf{G1: Unified MIS Abstraction.}
The system should provide a unified abstraction for diverse MIS types across different model architectures and training stages, enabling consistent extraction, storage, and downstream analysis.

\textbf{G2: Actionable Signal Aggregation.}
Given the fine-grained nature of MIS, the system should aggregate low-level signals into interpretable, decision-oriented summaries that can directly inform architectural actions, rather than exposing raw statistics.

\textbf{G3: Closed-Loop Optimization.}
The system should support a closed optimization loop that connects MIS analysis to concrete model modifications and evaluates their impact, enabling iterative refinement without manual intervention.

\textbf{G4: Low Exploration Cost.}
Compared to black-box AutoML, the system should significantly reduce the number of training or evaluation runs required to reach high-quality model configurations.

\textbf{G5: Generality and Interpretability.}
The framework should generalize across model families and optimization tasks, while producing human-interpretable signals that help engineers understand \emph{why} certain optimizations are effective.

\section{Related Work}
\label{sec:related_work}

Our work on MIS-driven optimization is situated at the intersection of black-box AutoML, model interpretability, and network pruning. AutoML and NAS methods automate model selection and architecture search, but they typically operate using only external performance metrics \cite{Feurer2015Efficient,zoph2017neural,liu2018darts,joglekar2019neuralinputsearchlarge,Real2020AutoML-Zero}. This black-box perspective leads to high exploration cost and limited interpretability, especially in industrial model optimization workflows.

A second line of work studies model explanations and feature attribution. Methods such as LIME, SHAP, and Integrated Gradients provide valuable insight into how models make predictions \cite{ribeiro2016lime,lundberg2017shap,sundararajan2017ig}. However, these methods are primarily designed to explain model behavior to humans rather than to drive automated model-edit decisions.

Finally, pruning and compression research demonstrates that internal signals can reveal under-utilized capacity and guide architecture refinement \cite{han2015pruning,yu2018nisp,frankle2019lottery}. MISO differs from these approaches by treating ranking, alignment, and comparison as reusable primitives inside a broader system for practical model optimization rather than a single-purpose compression technique.

\paragraph{Ranking Architectures.}
Our work is applied to ranking architectures that have evolved from early deep recommendation models~\cite{covington2016youtube,naumov2019dlrm} through increasingly modular designs such as DCN-V2~\cite{wang2021dcnv2} and stacked interaction architectures~\cite{zhang2022dhen}. These systems are characterized by heterogeneous feature types, large embedding tables, and frequent iteration cycles under tight latency budgets---properties that make undirected exploration expensive and MIS-guided optimization particularly attractive.

\begin{figure*}[!htbp]
  \centering
  \includegraphics[width=0.76\textwidth]{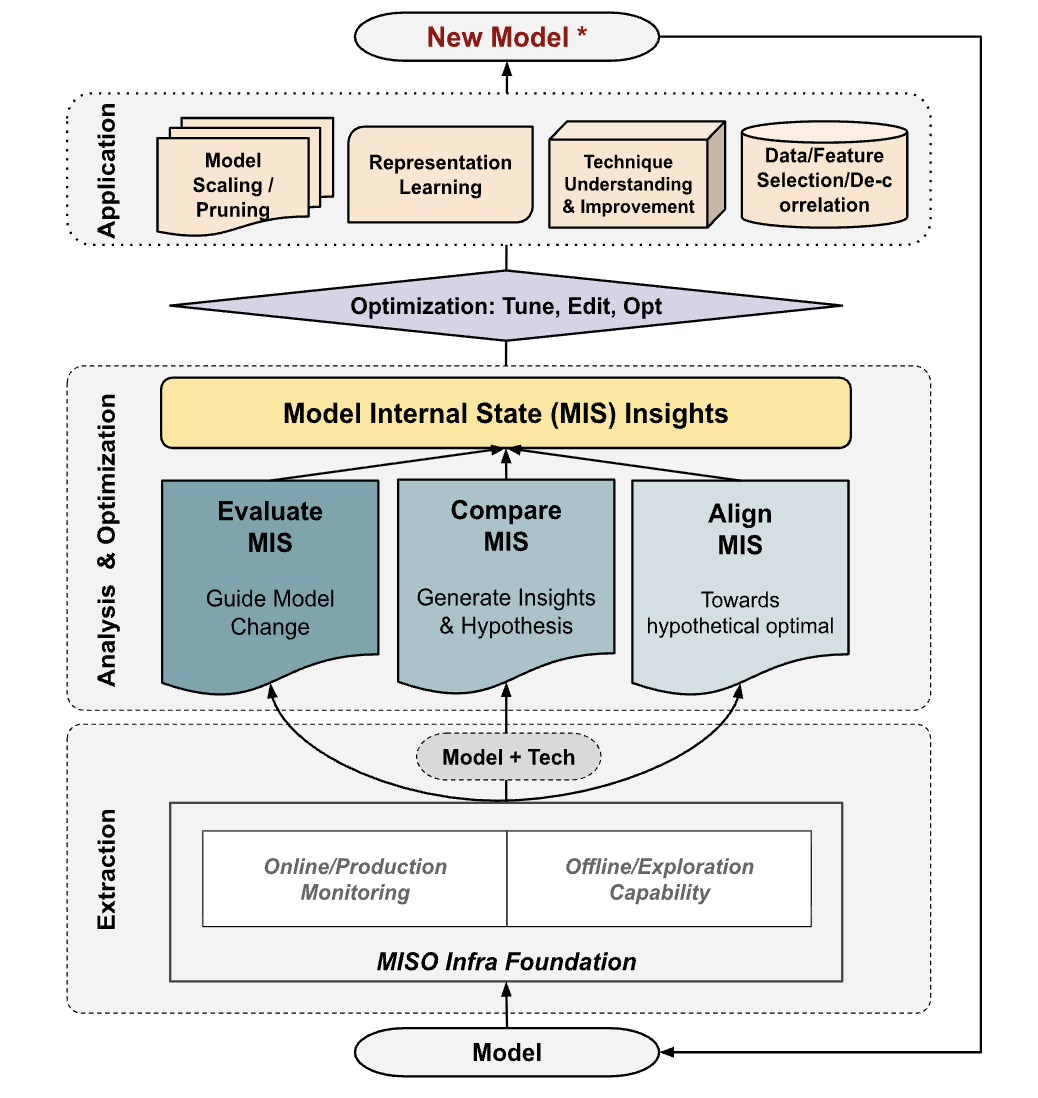}
  \caption{The MISO system architecture. The framework consists of three layers: (1) an \textbf{Extraction} layer that collects Model Internal States (MIS), (2) an \textbf{Analysis \& Optimization} layer that evaluates, compares, and aligns MIS to generate actionable insights, and (3) an \textbf{Application} layer where these insights drive concrete model optimizations such as scaling, pruning, representation learning, and feature selection.}
  \Description{Three-layer MISO architecture: an extraction layer collects model internal states; an analysis and optimization layer ranks, aligns, and compares those states; and an application layer uses the resulting insights for scaling, pruning, representation learning, and feature selection.}
  \label{fig:MISO}
\end{figure*}

\section{Architecture Overview}

MISO is designed as a general-purpose ML systems framework that closes the loop between model internal states (MIS) and model optimization actions. As shown in Figure~\ref{fig:MISO}, the system consists of three logical layers: (1) MIS Extraction Layer, (2) MIS Analysis and Aggregation Layer, and (3) Optimization and Decision Layer. Together, these layers transform fine-grained internal signals into actionable, low-cost model modifications.

\subsection{MIS Extraction Layer}

The MIS Extraction Layer provides a unified and extensible interface for collecting internal signals during offline training and post-training analysis. This layer abstracts away model- and framework-specific details and exposes a consistent representation of MIS, denoted as $\mathcal{S}(M)$.

MISO supports a wide range of MIS types, including model parameters, activations, gradients, normalization statistics, and neuron- or module-level metrics. Extraction can be configured at different granularities and time scales. By decoupling MIS extraction from downstream analysis, MISO enables the same optimization logic to be reused across different model architectures and workloads.

\subsection{MIS Analysis and Aggregation Layer}

Raw MIS are high-dimensional, heterogeneous, and often too fine-grained to directly inform optimization decisions. The MIS Analysis and Aggregation Layer bridges this abstraction gap by transforming $\mathcal{S}(M)$ into \emph{actionable summaries} through a set of MIS aggregation primitives.

Each aggregation primitive implements a mapping $f: \mathcal{S}(M) \rightarrow \mathcal{A}$, where $\mathcal{A}$ represents decision-oriented signals such as rankings, scores, or constraints. We instantiate three classes of primitives.

\textbf{Ranking-based primitives} summarize MIS into importance or ROI scores. A representative instantiation is neuron importance, where we combine gradient-based and perturbation-based signals:
\begin{equation}
I_n = \alpha \cdot \frac{1}{|\mathcal{D}|} \sum_{x \in \mathcal{D}} |a_n \cdot \nabla_{a_n} \mathcal{L}(x)| + (1-\alpha) \cdot \Delta \mathcal{L}|_{a_n \leftarrow \text{shuffle}(a_n)}
\label{eq:neuron_importance}
\end{equation}
where $\mathcal{L}$ is the loss, $\mathcal{D}$ is validation data, and $\alpha$ is a balancing hyperparameter.

\textbf{Alignment-based primitives} diagnose whether intermediate distributions deviate from a layer-specific reference. For normalized layers, a useful reference is the intended standardized distribution. For a given layer $l$, we compute the following diagnostic score:
\begin{equation}
\text{Align}(l) = 1 - \frac{D_{KL}(p_{pre}^l || \mathcal{N}(0,1)) + D_{KL}(p_{post}^l || \mathcal{N}(0,1))}{2}
\label{eq:alignment_score}
\end{equation}
Here, $p_{pre}^l$ and $p_{post}^l$ denote empirical pre- and post-normalization distributions. The score is used to rank layers for inspection rather than as a universal measure of representation quality. \textbf{Comparison-based primitives} contrast MIS across models, layers, or training stages to surface anomalies and identify where candidate variants diverge.

\subsection{Optimization and Decision Layer}

The Optimization and Decision Layer translates aggregated MIS signals into concrete model modification actions. Actions are defined over a configurable action space, including parameter scaling, pruning, module replacement, and feature reconfiguration.

MISO implements a closed-loop workflow. Given an initial model, MIS are extracted and aggregated to produce actionable signals, which are then mapped to a small set of candidate modifications. These candidates are evaluated under budget constraints, and the resulting feedback is used to refine the next round of decisions.

\section{MIS Instantiations in Ranking Models}

We instantiate MISO on large-scale ranking models to demonstrate how the framework maps model internal states to optimization actions in a real-world workload. Ranking models present a representative and challenging setting due to their heterogeneous architectures, large parameter counts, and stringent efficiency constraints.

\subsection{Target Model and Optimization Scope}

\begin{figure}[!t]
  \centering
  \includegraphics[width=0.82\columnwidth]{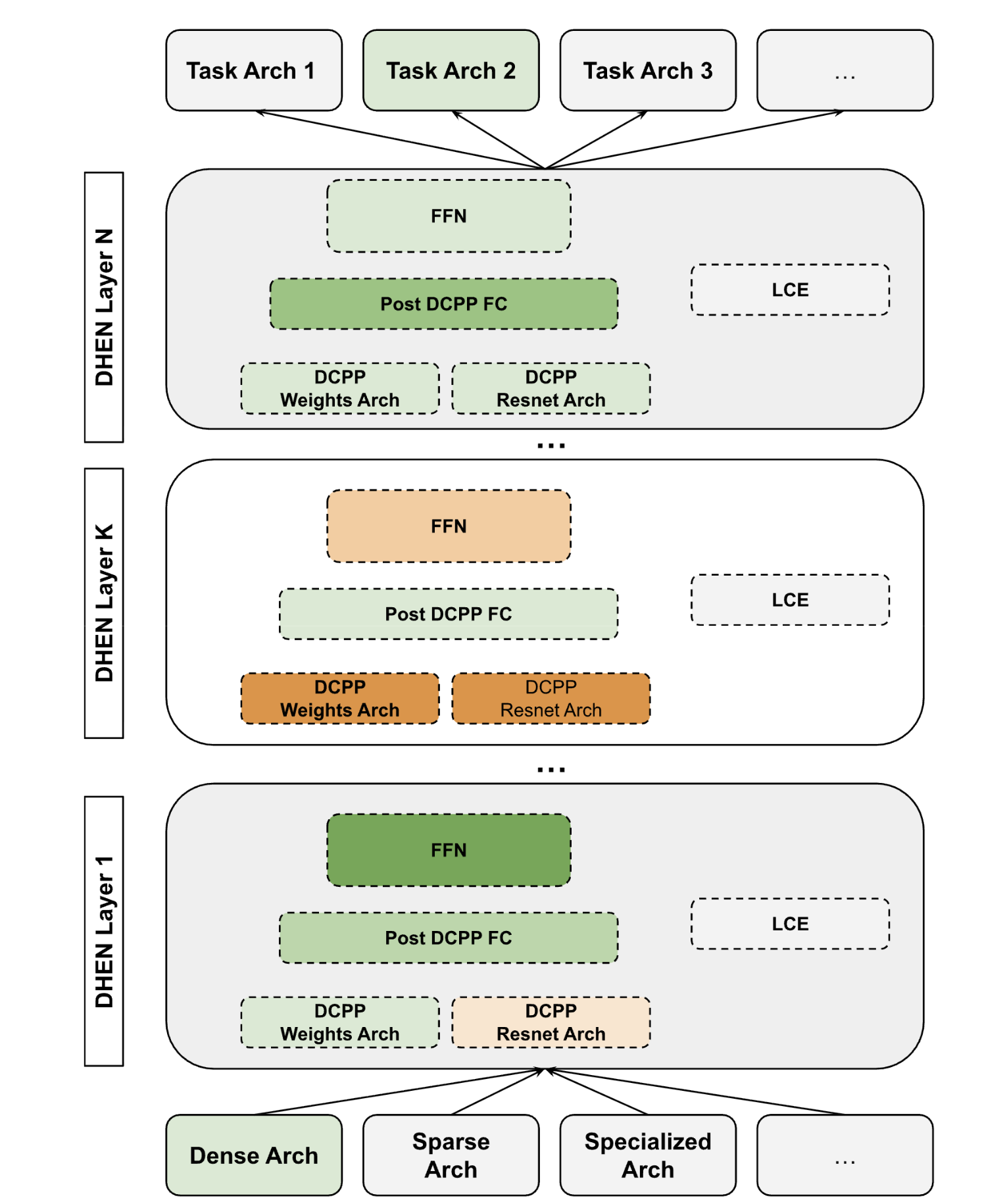}
  \caption{Representative ranking model architecture used to instantiate MISO. The model combines multiple feature streams and stacked submodules, creating several natural decision points for MIS-guided scaling, replacement, and refinement.}
  \Description{A ranking-model architecture with multiple task heads, stacked DHE layers containing feed-forward and DC/PP components, and several dense, sparse, and specialized architecture choices.}
  \label{fig:model_arch}
\end{figure}

The target workload consists of deep neural ranking models used in recommendation and retrieval systems. These models combine embedding layers, feature-interaction modules, multi-layer perceptrons, and task-specific heads, resulting in tens to hundreds of millions of parameters. Figure~\ref{fig:model_arch} shows a representative architecture used in our study. Within this setting, MISO focuses on optimization tasks that are common yet costly in practice: capacity scaling, parameter pruning, architectural refinement, and module replacement.

The optimization objectives include model quality metrics such as normalized entropy together with system-level constraints such as computational efficiency and training cost. Importantly, MISO operates under limited exploration budgets, where exhaustive search or brute-force AutoML is infeasible. This makes ranking models a suitable stress test for MIS-guided optimization: the search space is large enough to make undirected exploration expensive, yet the engineering workflow still expects interpretable, modular decisions.

\subsection{Collected MIS Types and Action Mapping}

To support these optimization tasks, MISO instantiates MIS extraction for a representative set of internal signals in ranking models. Specifically, MISO collects: (1) neuron- and module-level parameter statistics, including weights and gradients; (2) activation statistics at key interaction and transformation layers; (3) normalization-related signals such as mean and variance before and after normalization layers; and (4) feature- and module-level response statistics that reflect sensitivity to perturbations.

These signals are mapped to optimization actions through the aggregation primitives. Ranking-based MIS are used primarily for selective scaling and pruning; alignment-based MIS are used to identify unstable normalization or representation modules; and comparison-based MIS are used to explain performance differences across nearby variants and support architectural refinement.

\section{Experimental Setup}

We evaluate MISO on large-scale ranking models as a systems case study. We ask whether MIS-guided proposals improve ranking quality under a bounded validation budget, whether they reduce the number of full training runs, and how the three MIS primitives contribute to the resulting recommendations.

\subsection{Workloads, Baselines, and Metrics}

The evaluation uses recommendation-style ranking workloads \cite{covington2016youtube,naumov2019dlrm}. The dataset comprises billions of training examples with both dense and sparse features, representative of ads recommendation workloads. We compare MISO with two practical baselines. \textbf{Black-box scaling} expands model capacity without MIS guidance. \textbf{Expert-driven tuning} reflects a standard workflow in which engineers choose the next experiment from end-to-end metrics, domain knowledge, and available diagnostic signals.

We study three recurring optimization tasks: capacity scaling based on neuron or module importance, replacing poorly aligned normalization blocks, and refining architectural choices using cross-model comparison signals. We report Normalized Entropy (NE) improvement relative to black-box scaling together with the number of training runs required to reach a satisfactory configuration. Normalized Entropy (NE) measures the predictive calibration of a ranking model relative to the empirical base rate; lower NE indicates better ranking quality. 

\subsection{Evaluation Protocol and Fairness}

Our goal is not to ask which method can discover the best model after unlimited search, but which workflow reaches a useful proposal under a realistic engineering budget. We therefore report both final quality and the number of training runs required to reach the selected configuration. This framing reflects the practical setting in which retraining and validation dominate the cost of local model refinement.

MISO incurs additional analysis work to extract MIS, but all methods are evaluated against the same downstream objective and deployment constraints. We do not claim that internal-state analysis is free. Instead, the case study evaluates whether this up-front analysis is a favorable tradeoff when the dominant cost is the number of full model trials. The reported run counts should therefore be interpreted as an engineering-efficiency measure, not as a claim that MISO eliminates all optimization cost.

\subsection{Representative Optimization Cases}

MISO is designed to support several recurring optimization patterns rather than a single fixed heuristic. In our ranking workloads, ranking-based signals are most useful for selective scaling, because they identify where extra capacity is likely to pay off. Alignment-based signals are most useful when the issue is unstable intermediate statistics rather than raw capacity, motivating normalization-module replacement or related refinements. Comparison-based signals become valuable when nearby variants differ subtly and engineers need to isolate where their internal behavior diverges.

These three cases illustrate why MIS are useful in practice. The system is not replacing engineering judgment with an opaque score. Instead, it narrows the decision space using evidence that remains interpretable at the module or architectural level.

\section{Results}

\subsection{Main Results}

Table~\ref{tab:main_results} summarizes the quality and engineering-efficiency results. Across the evaluated model scales, MISO yields larger relative NE improvements than expert-driven tuning while using substantially fewer validation runs. At the 50x scale, the relative NE improvement obtained with MISO is $2\times$ that obtained with expert-driven tuning.

These gains are accompanied by a large reduction in the number of evaluated configurations. Across the reported scales, MISO requires 3--12 training runs versus 50--92 for expert-driven tuning, a reduction of 84--94\%. In this case study, internal-state analysis is useful not only for selecting a better configuration but also for reducing the search budget required to reach it.

\begin{table}[!t]
\centering
\small
\caption{Main results across model scales. NE improvements are reported as relative ratios over the expert-driven baseline.}
\label{tab:main_results}
\resizebox{\columnwidth}{!}{%
\begin{tabular}{lcccc}
\toprule
\textbf{Metric} & \textbf{13x} & \textbf{20x} & \textbf{32x} & \textbf{50x} \\
\midrule
Relative NE improvement (MISO vs.\ expert-driven) & $2.5\times$ & $2.0\times$ & $2.0\times$ & $2.0\times$ \\
Expert-driven runs & 50$\pm$10 & 62$\pm$12 & 50$\pm$10 & 92$\pm$18 \\
MISO runs & \textbf{3$\pm$1} & \textbf{5$\pm$2} & \textbf{8$\pm$2} & \textbf{12$\pm$3} \\
Run reduction & 94\% & 92\% & 84\% & 87\% \\
\bottomrule
\end{tabular}}
\end{table}

\begin{figure*}[!tb]
  \centering
  \makebox[\textwidth][c]{%
  \begin{minipage}[t]{0.42\textwidth}
    \centering
    \includegraphics[width=\linewidth]{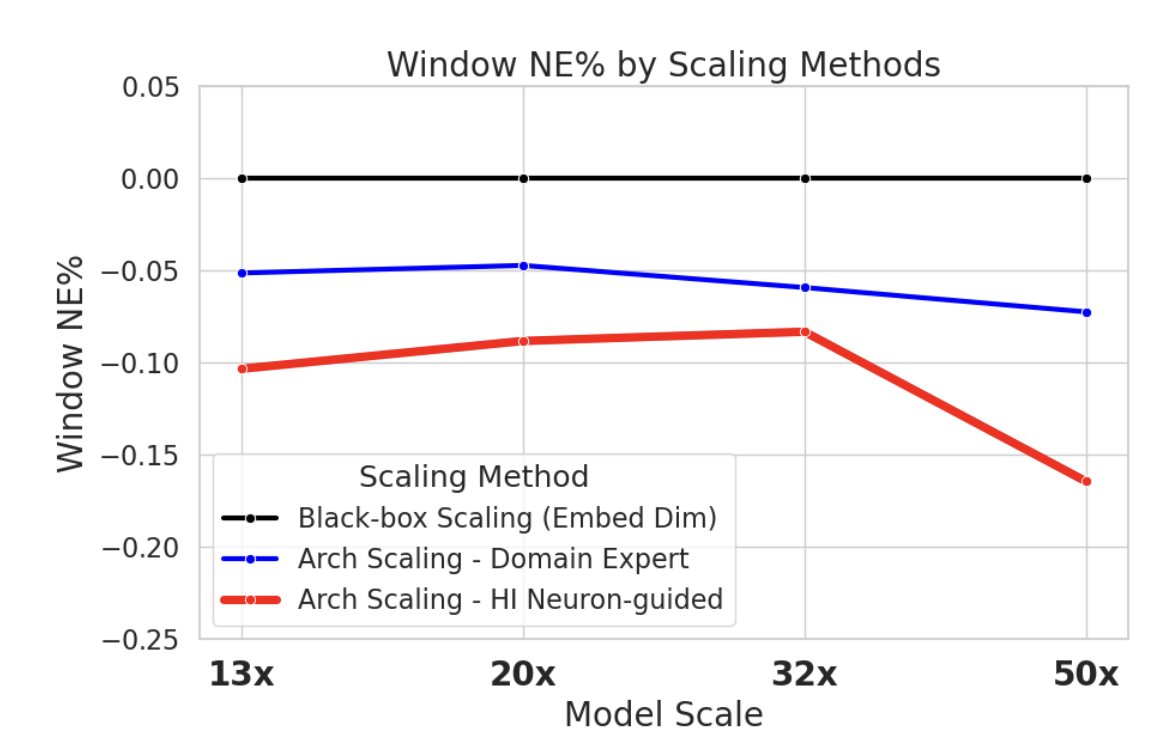}
  \end{minipage}\hfill
  \begin{minipage}[t]{0.42\textwidth}
    \centering
    \includegraphics[width=\linewidth]{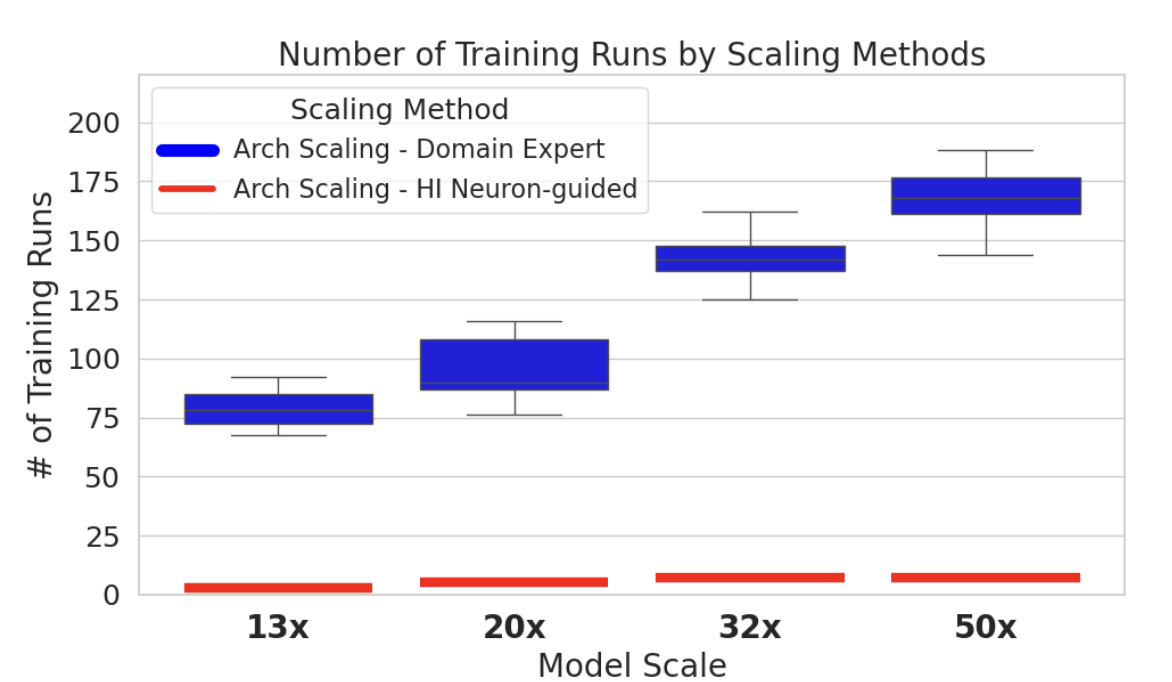}
  \end{minipage}%
  }
  \caption{Scaling results across model sizes. \textbf{Left:} NE improvement relative to black-box scaling. \textbf{Right:} Training runs required. MISO consistently improves quality while sharply reducing exploration cost.}
  \Description{Two bar charts across four model scales. The left chart shows that MISO yields greater normalized-entropy improvement than black-box scaling. The right chart shows that MISO requires substantially fewer training runs.}
  \label{fig:scaling_results}
\end{figure*}

\subsection{Contribution of Different MIS Primitives}

Different MIS types support different optimization actions. Ranking-based signals are most useful for selective scaling. Alignment-based signals are most helpful when the issue is unstable normalization or representation statistics. Comparison-based signals become valuable when deciding between nearby architectural variants and isolating the source of performance differences.

This pattern appears in the ablation study at the 50x scale shown in Table~\ref{tab:ablation}. Using ranking information alone already improves quality, but adding alignment and comparison signals yields the strongest end-to-end recommendations and modest efficiency gains in model size and latency. The result is not that any one primitive is universally sufficient, but that the combination forms a reusable interface for several common optimization actions.

\begin{table}[!t]
\centering
\caption{Ablation study on MIS primitives (50x scale). Relative NE gain is normalized to the full MISO configuration (=100\%).}
\label{tab:ablation}
\resizebox{\columnwidth}{!}{%
\begin{tabular}{lccc}
  \toprule
  Configuration & Relative NE gain & Params & Latency \\
  \midrule
  Baseline                 & 0\%    & 100\% & 100\% \\
  + Ranking only           & 62\%   & 100\% & 100\% \\
  + Ranking + Alignment    & 62\%   & 100\% & 92\%  \\
  + Ranking + Comparison   & 75\%   & 95\%  & 92\%  \\
  Full MISO                & 100\%  & 95\%  & 92\%  \\
  \bottomrule
  \end{tabular}}
\end{table}

\subsection{Takeaways}

From a systems perspective, the key result is that internal-state analysis can shrink the validation budget for local model optimization without turning the process into an opaque search loop. MISO does not attempt to replace all of AutoML or solve general NAS. Instead, it offers a practical middle ground between standard expert workflows and black-box exploration: engineers choose the optimization scope, while MISO supplies ranked, interpretable proposals that reduce wasted trials.

This positioning is especially relevant for ranking and recommendation systems, where absolute metric gains are often incremental but can still justify deployment if the path to those gains is inexpensive and interpretable. In such settings, saving dozens of training runs can be as important as the final performance delta itself.

\section{Discussion}

\subsection{Operational Interpretation}

MISO is most useful in settings where the model family is already fixed and the real problem is repeated local optimization. In such workflows, engineers are rarely deciding among radically different architectures; instead, they are deciding where to add capacity, which module to replace, or which nearby variant is worth another expensive run. MISO matches this decision regime well because it produces recommendations at the same granularity as these engineering actions.

The framework is also attractive because it preserves interpretability. A recommendation produced by MISO is accompanied by a concrete internal signal pattern---for example, under-utilized neurons, poorly aligned normalization statistics, or a divergence between two variants at a specific layer. This is operationally useful even when the final metric gain is modest, because it provides a rationale that can be audited by practitioners.

\subsection{When MISO Helps Less}

MISO is less helpful in settings where a model has not yet reached a usable baseline, where internal signals are too noisy to be stable, or where the dominant question is broad architecture discovery rather than targeted refinement. In these regimes, black-box exploration or other search procedures may still be necessary. The intended contribution of MISO is therefore not to replace all forms of AutoML, but to introduce a practical optimization layer for an important class of repeated engineering decisions.

\section{Limitations and Future Work}

MISO should be read as a practical optimization system rather than a claim of universal algorithmic superiority. First, the evaluation is a ranking-model case study, and broader validation across domains and model families remains future work. Second, comprehensive MIS extraction introduces analysis overhead, especially for perturbation-based signals. Third, the current evidence concerns post-hoc refinement of trained models rather than cold-start search over entirely new architectures. Finally, the reported results are aggregate operational measurements from a proprietary deployment setting; they should not be interpreted as a fully reproducible benchmark or as a statistical claim about every ranking workload.

Future work includes extending the framework to additional model families, reducing the cost of MIS extraction, and using language-model-based agents to assist the interpretation of internal-state patterns.

\subsection*{Reproducibility}
Due to deployment constraints on the underlying models and data, we cannot release the full codebase; the paper is therefore positioned as a systems and application case study, with algorithms and decision workflow described in enough detail to support adaptation on similar workloads.

\section{Conclusion}

We presented \textbf{MISO}, a system for model-internal-state (MIS)-guided refinement of ads ranking models. MISO extracts and aggregates diverse internal signals—including neuron importance, activation alignment, and cross-model comparison—to guide model modifications in an interpretable manner.

Across the evaluated ranking-model scales, MISO improves normalized entropy while requiring only a small fraction of the training trials used by expert-driven tuning. More broadly, the case study suggests that model internal states can serve as a practical optimization interface for repeated real-world ranking-model refinement.

\bibliographystyle{ACM-Reference-Format}
\bibliography{bibtex}

\end{document}